\documentclass[11pt,letterpaper,usenames,dvipsnames]{article}
\usepackage{jheppub}

\usepackage{enumitem}

\RequirePackage{pdf14}
\usepackage[utf8]{inputenc}
\usepackage{graphicx,tikz}
\usepackage[framemethod=TikZ]{mdframed} % package used to make the fancy boxes
\usepackage{amsfonts,amsmath,amssymb,amsthm,esint}
\usepackage{array,setspace,mathrsfs,yfonts,dsfont,bbm,colonequals,amscd,mathtools}
\usepackage{relsize,suffix,cancel,bbm,capt-of,upgreek,verbatim,xcolor}
\usepackage[T1]{fontenc}
\usepackage{enumitem}
\usepackage{multirow}
\usepackage{arydshln}
\usepackage{arydshln}
\usepackage{tcolorbox}

\usepackage{tikz} %figures using tikz package
\usetikzlibrary{arrows,snakes,shapes.arrows,decorations.markings}
     \tikzset{>=triangle 90}
     \tikzstyle{gr}=[draw,circle,green!50!black,fill=green!50!black,scale=.6]
     \tikzstyle{Bl}=[draw,circle,blue,scale=.6]
     \tikzstyle{R}=[draw,circle,fill=red,scale=.6]
     \tikzstyle{bl}=[draw,circle,fill=black,scale=.35]
     \tikzstyle{bbc}=[draw,circle,fill=black,scale=.75]
     \tikzstyle{bbcs}=[draw,circle,fill=black,scale=.5]
     \tikzstyle{rc}=[circle,fill=red,scale=.6]
     \tikzstyle{wc}=[draw,circle,scale=.75]

\def\del{{\partial}}

\def\bar{\overline}

\def\^{\wedge}
\def\I{\mathds{1}}

\def\Im{\mathop{\rm Im}}

\def\U{{\rm U}}
\def\SU{{\rm SU}}

\def\Sp{\mathop{\rm sp}}

\def\br{{\bf r}}

\def\a{{\alpha}}

\def\b{{\beta}}
\def\g{{\gamma}}

\def\D{{\Delta}}

\def\l{{\lambda}}
\def\L{{\Lambda}}

\def\s{{\sigma}}

\def\S{{\Sigma}}
\def\t{{\tau}}
\def\ch{{\chi}}

\def\w{{\omega}}

\def\ff{{\mathfrak f}}

\def\nf{{\mathfrak n}}

\def\spf{\mathfrak{sp}}
\def\suf{\mathfrak{su}}

\def\uf{\mathfrak{u}}

\def\mP{\mathsf{P}}

\def\cA{{\mathcal A}}
\def\cB{{\mathcal B}}

\def\cC{{\mathcal C}}
\def\cCrg{{\mathcal C}_{\rm reg}}

\def\cG{{\mathcal G}}

\def\cK{{\mathcal K}}
\def\cL{{\mathcal L}}

\def\cN{{\mathcal N}}

\def\cS{{\mathcal S}}
\def\cSb{{\bar{\cS}}}
\def\cSbme{{\bar{\cS}}_{\rm metr}}
\def\cSbcp{{\bar{\cS}}_{\rm cplx}}
\def\cT{{\mathcal T}}
\def\cTu{{\mathcal T}_\bu}

\def\C{\mathbb{C}} 
\def\DD{\mathbb{D}}

\def\P{\mathbb{P}}
  
\def\R{\mathbb{R}}

\def\Z{\mathbb{Z}}

\def\Csc{\mathscr{C}}

\def\Psc{\mathscr{P}}

\def\beq{\begin{equation}}
\def\eeq{\end{equation}}
\def\bea{\begin{align}}
\def\eea{\end{eqnarray}}

\newcommand{\bpmat}{\begin{pmatrix}}
\newcommand{\epmat}{\end{pmatrix}}
\newcommand{\bsmat}{\begin{smallmatrix}}
\newcommand{\esmat}{\end{smallmatrix}}

\def\Dph{D^{\rm phys}_x}

\newtheorem{fact}{Fact}

\def\ba{{\bf a}}

\def\bh{{\boldsymbol h}}

\def\bm{{\bf m}}
\def\bn{{\bf n}}

\def\bu{{\boldsymbol u}}

\def\onesl{\textsl{1}}
\def\twosl{\textsl{2}}
\def\threesl{\textsl{3}}

\begin{document}

\title{Towards the classification of rank-$r$ $\mathcal{N}=2$ SCFTs. Part I: twisted partition function and central charge formulae.}

\author[1]{Mario Martone}

\affiliation[1]{University of Texas, Austin, Physics Department, Austin TX 78712}

\emailAdd{mariomartone@utexas.edu}

\abstract{We derive explicit formulae to compute the $a$ and $c$ central charges of four dimensional $\cN=2$ superconformal field theories (SCFTs) directly from Coulomb branch related quantities. The formulae apply at arbitrary rank. We also discover general properties of the low-energy limit behavior of the flavor symmetry of $\cN=2$ SCFTs which culminate with our $\cN=2$ \emph{UV-IR simple flavor condition}. This is done by determining precisely the relation between the integrand of the partition function of the topologically twisted version of the 4d $\cN=2$ SCFTs and the singular locus of their Coulomb branches. The techniques developed here are extensively applied to many rank-2 SCFTs, including new ones, in a companion paper. %We believe that the results presented here and in \cite{Argyres:2020b}, represent an important qualitative step forward in extending the classification of four dimensional $\cN=2$ SCFTs beyond rank-1.
\\\vspace{3 cm}

This manuscript is dedicated to the memory of Rayshard Brooks, George Floyd, Breonna Taylor and the countless black lives taken by US police forces and still awaiting justice. Our hearts are with our colleagues of color who suffer daily the consequences of this racist world.
}
\maketitle
\setcounter{page}{1}

\section{Introduction}

Since Witten's seminal paper \cite{Witten:1988ze}, the connection between topological and supersymmetric field theories has been fruitful. Beyond the deep insights that it brought into the study of four manifolds \cite{Witten:1994cg,Witten:1994ev,Donaldson:1996}, see \cite{Dedushenko:2017tdw,Gukov:2017zao,Moore:2017cmm} and references therein for recent developments,  the fact that supersymmetric theories contain a sub-sector which is topological has brought considerable progress to our understanding of supersymmetric field theories themselves, see e.g. \cite{Nekrasov:2002qd,Nekrasov:2003rj,Shapere:2008zf,Pestun:2007rz}. The procedure of projecting onto the topological sector of a supersymmetric field theory is now standard and known as topological twist. The work presented here represents yet another instance in which the constrained picture of topological field theories is used to advance our understanding of the properties of the their non-topological cousins.

Our interest in studying topologically twisted $\cN=2$ theories, and in particular their partition functions, is motivated by our ambitious classification program of $\cN=2$ superconformal field theories (SCFTs) in four dimensions. This program, see \cite{Argyres:2020nrr} for a lightning review and a list of relevant references, is based on a systematic study of Coulomb branches whose geometry appears to contain enough structure to nearly determine the space of allowed $\cN=2$ SCFTs.\footnote{The qualification ``nearly'' refers to possible ambiguities which arise if we also allowed for discrete gauging \cite{Argyres:2016yzz}.} The results of this paper give further evidence for this fact as we are able to derive general formulae which allow to straightforwardly compute, directly from Coulomb branch related quantities, the $(a,c)$ central charges, as well as the flavor levels $k_\ff$, of any $\cN=2$ SCFT at arbitrary rank.

It is useful to summarize the main results of this paper right away:
\begin{itemize}

\item[$i$.] We derive explicit formulae which express in terms of Coulomb branch quantities the $(a,c)$ central charges and the level of any simple flavor factor $\ff$, of an arbitrary $\cN=2$ SCFT :
\begin{subequations}
\begin{align}
\label{actotaint}
24 a &= 5r + h + 6 \left(\sum_{\ell=1}^r\D_{\bu_\ell}-r\right) +\sum_{i\in I}\D^{\rm sing}_i \frac{12 c_i-2-h_i}{\D_i} ,
\\\label{actotbint}
12 c &= 2r + h + \sum_{i\in I}\D^{\rm sing}_{i} \frac{12 c_i-2-h_i}{\D_i},\\\label{actotcint}
k_\ff&=\sum_{i\in I_{\ff}}\frac{\D_i^{\rm sing}}{d_i\D_i} \left(k^i-T({\bf2}\bh_i)\right)+T({\bf2}\bh).
\end{align}
\end{subequations}
Here $r$ is the rank of the SCFT, $h$ is the quaternionic dimension of the theory's extended Coulomb branch, $\D_{\bu_\ell}$ is the scaling dimension of the theory's $\ell$-th Coulomb branch  parameters, $\D^{\rm sing}_i$ is defined in \eqref{Dsing} and all the remaining quantities indexed by $i$ refer to known quantities of rank-1 theories which are supported on special loci of the Coulomb branch (see section \ref{sec:lowene} below). We call these formulae \emph{central charge formulae}.

The central charge formulae, which systematize and extend the  beautiful work by Shapere and Tachikawa \cite{Shapere:2008zf}, are derived in section \ref{sec:Twisted} and \ref{sec:flavor} using the well-known relation between the $(a,c,k_{\ff})$ central charges of an untwisted $\cN=2$ SCFT and the $\U(1)_R$ symmetry anomaly of its twisted topological version.%, we use the  translate the understanding in $i.$ to the general formulae, \eqref{actota}-\eqref{actotb}-\eqref{actotc}, which relate the $(a,c,k_{\ff})$ of any rank-$r$ SCFTs, to well-known quantities of rank one theories and which can be computed in the low-energy effective theory.

\item[$ii$.] We formulate and prove the $\cN=2$ \emph{UV-IR simple flavor condition}. This condition states that all mass deformations of a rank $\ge2$ $\cN=2$ SCFT modify the asymptotics of the CB and are realized as mass deformations of the same rank-1 theories which appear in their central charge formulae. This realization de-facto reduces the problem of analyzing mass deformations of higher rank theories to that of rank-1,\footnote{Note that this result only applies to mass deformations, and not to the other class of relevant $\cN=2$ deformations known as chiral deformations.} providing a clear route towards a systematic understanding of the space of higher rank theories.

The UV-IR simple flavor condition is discussed in section \ref{sec:flavor} and it is a straightforward consequence of the structure of the integrand of the partition function of the twisted version of the $\cN=2$ SCFT.

\item[$iii$.] We determine the relation between the integrand of the partition function of a topologically twisted $\cN=2$ SCFT and the structure of the singular locus of its Coulomb branch at arbitrary rank. This precise relation is used copiously to derive the results in $i.$ and $ii.$ and lead to a sharp definition and understanding of both the \emph{quantum} and \emph{physical} discriminant, see section \ref{sec:QuaDis} and \ref{sec:PhyDis}.

\end{itemize}

It is, of course, not the first time that the partition function of topologically twisted $\cN=2$ theories with rank higher than one is studied in detail. The fact that the zeros and the singularities of the measure of the partition function could only lie along the singular locus, which is used here as a starting point, was noticed since the early days \cite{Witten:1995gf,Losev:1997tp,Moore:1997pc,Marino:1997gj,Marino:1998bm,Marino:1998tb}. Here we point out that the orders of such zeros is subtle to determine but we are able to carry out this calculation in general. This is not just a mathematical curiosity. In fact our formulae above, precisely rely on a detailed determination of these exponents.

The paper is organized as follows. In the next section we will give a reminder of the geometric structure of Coulomb branches of $\cN=2$ theories at arbitrary rank and discuss the singular locus and the concept of the discriminant in detail. In section \ref{sec:Twisted} we will set up the study of the partition function of a topologically twisted $\cN=2$ SCFT and derive our central charge formulae. Section \ref{sec:flavor} will instead be dedicated to a systematic study of the flavor structure of $\cN=2$ SCFTs and making many general remarks connecting the behavior of the flavor symmetry at high and low energies. We end the paper with our conclusions.
\section{Discriminant locus, quantum discriminant, and physical discriminant}\label{sec:2}

The central charge formulae \eqref{actotaint}-\eqref{actotcint} depend on a series quantities indexed by $i\in I$ which are associated to rank-1 theories appearing on special loci of the Coulomb branch (CB). We define these quantities here and explain our definitions in explicit examples. In doing so we clarify what we mean by discriminant of the CB a term which is often used in the literature in different contexts and with different meanings. Specifically we will give our definitions for three different but related objects: the \emph{discriminant locus}, the \emph{quantum discriminant}, or loosely the discriminant of the Seiberg-Witten (SW) curve, and the \emph{physical discriminant} which instead enters the measure factor of the partition function of the twisted $\cN=2$ theory. These definitions will be needed in the next section. To understand this discussion, it might be useful to consult appendix \ref{app:CB} where the basics of CB geometry and elements of the \emph{special Kahler stratification} are given.

%the geometry described by $\S$ is scale invariant and a $\C^*$-action, which physically arises \red{from} combining the $\R^+$ dilatation and the $\U(1)_R$ action, is defined over both its base and its fiber.\footnote{\red{Why should it be possible to choose the SW fiber to have a $\C^*$ action?  I think it's true in the rank 1 and 2 cases, but am not sure about higher ranks... Do you really need the $\C^*$ action on the fiber anywhere?}}

%Our main objects of study in this paper are $\cN=2$ superconformal field theories.  This superconformal  

\subsection{Structure of the singularity}

The low-energy theory on a generic point of the CB $\cC$ is almost as boring as it gets; a free $\cN=2$ supersymmetric $\U(1)^r$ gauge theory with no massless charged states. $r$ is called the \emph{rank} of the theory and coincides with the complex dimensionality of $\cC$, ${\rm dim}_\C\cC=r$. $\cC$ is a singular space and its singular locus will be at the center of our discussion and denoted as $\cSb$. $\cSb$ stands both for ``singular'' and ``stratum'', the reason for the bar is that $\cSb$ is a closed subset of $\cC$ and the smooth part of the CB is $\cCrg := \cC \setminus \cSb$.  Thus $\cCrg$ is an open subset of $\cC$. 

Henceforth we will only discuss the special case when the $\cN=2$ theory is superconformal in which case the symmetry group includes an $\R^+ \times \U(1)_R$ (we are neglecting the $\SU(2)_R$ factor as it acts trivially on $\cC$) which can be spontaneously broken, and so acts nontrivially on $\cC$, and combines to give a $\C^*$ action on the CB. We will label the $\C^*$ weight or scaling dimension of a quantity $\cdot$ interchangeably as $[\cdot]$ or $\D(\cdot)$. The origin of $\cC$ is the single vacuum of the entire moduli space which is invariant under such action and for this reason we will call it the superconformal vacuum. The entire structure of $\cC$ has to be compatible with the $\C^*$ action and in particular $\cSb$ and $\cCrg$ have to be closed under it. We will often refer to the set of constraints arising from the compatibility with the $\C^*$ action as the constraints coming from scale invariance. In this language we will only consider here scale invariant CB geometries.

Finally, the CB is both a complex space and a metric space, and so $\cC$ can have singularities in each of these structures \cite{Argyres:2018wxu} and the singular locus is the union of the loci of the two types of singularities, $\cSb = \cSbme \cup \cSbcp$. The physics interpretations of the two are remarkably different.  $\cSbcp$ is the locus of vacua for which the operators generating the corresponding CB chiral ring satisfy non-trivial relations.  This means that the chiral ring is not freely generated at points in $\cSbcp$.  In the following we will make the assumption that $\cSbcp=\varnothing$, therefore $\cSb\equiv\cSbme$. This condition can be equivalently stated as assuming that $\cC$ admits a set of globally defined $r$ complex coordinates which have definite weight under the $\C^*$ action and which we will collectively indicate as $\bu$. 

\subsection{Discriminant locus}

$\cSbme$ is the locus of $\cC$ where extra charged states become massless or, in other words, where the low-energy physics is not captured solely by a bunch of free $\cN=2$ vector multiplets. The central charge $|Z_Q|$, see \eqref{ccBPS}, is a lower bound on the mass of a state with charge $Q$,  therefore $Z_Q(\bu)$ vanishes for any $\bu\in\cSbme$. Assuming away some pathological behavior and carefully keeping track of the structure of the CB geometry, it is possible to prove that $\cSb$ is an $r-1$ complex dimensional algebraic subvariety of $\cC$, which is the union of connected, irreducible, components $\cSb_i$:
\beq\label{stra1}
\cSb := \bigcup_{i\in I}\cSb_i,\qquad 
\cSb_i := \Bigl\{\bu\in\cC \,\Big|\, Z_Q\left(\s(\bu)\right)=0, \quad \forall \, Q\in\L_i \Bigr\}.
\eeq
Each component $\cSb_i$ is defined by the vanishing of the central charge for charges corresponding to the set of BPS states in the theory which become massless there. Since $Z_Q$ is linear in the charges, what defines $\cSb_i$ is the sublattice of charges $\L_i$ given by the integer linear span of such charges. By charge conservation, $\L_i$ is constant along $\cSb_i$ irregardless of walls of marginal stability. The index $i$ runs over some finite set $I$ which corresponds to the range of the sum of the equations we have reported in the introduction \eqref{actotaint}-\eqref{actotcint}. We will discuss in the next section how a rank-1 theory can be associated to each component $\cSb_i$ % Also each $\cSb_i$  $\L_i$ is at most a rank-2 sublattice of the full lattice of charges of the superconformal theory \cite{Argyres:2018zay}.

Since \eqref{stra1} is a complex co-dimension one algebraic subvariety of $\cC$ it can be cut out by a single polynomial on the CB, which is a product of polynomials whose zero locus corresponds to each component $\cSb_i$.  If this polynomial is reduced, then it is unique up to an overall constant factor. We then define the \emph{discriminant locus} to be the following quantity:
\beq\label{DisLoc}
D_x:=\prod_{i\in I} P_i(\bu),\qquad \cSb_i=:\Big\{\bu\in\cC\Big| P_i(\bu)=0\Big\}
\eeq
with $\cSb_i$ as in \eqref{stra1} and the $P_i(\bu)$ are distinct and irreducible for all $i\in I$. By scale invariance, the $P_i(\bu)$ are weighted homogeneous polynomials in the $\bu$, and we will call $\D^{\rm sing}_i$ the weight of $P_i(\bu)$ under the $\C^*$ action:
\beq\label{Dsing}
\D_i^{\rm sing}:=\D\Big(P_i(\bu)\Big).
\eeq
This quantity plays an important role in our central charge formulae \eqref{actotaint}-\eqref{actotcint}.

\begin{center}
\rule[1mm]{2cm}{.4pt}\hspace{1cm}$\circ$\hspace{1cm} \rule[1mm]{2cm}{.4pt}
\end{center}

\paragraph{Example: SCFTs with $\suf(3)$ gauge algebra} It might be useful to immediately follow this definition with a concrete example. Consider an $\suf(3)$ gauge theory with hypermultiplets in representations for which the beta function for the gauge coupling vanishes. It is a straightforward group theory exercise to show that there are three possibilities:
\beq\label{theories}
\suf(3)\ \cN=2\ {\rm SCFTs}:\left\{
\begin{array}{cl}
\textsl{1}.&\ 6({\bf 3})\\
\textsl{2}.&\ 1({\bf 6})\oplus 1 ({\bf 3})\\
\textsl{3}.&\ 1({\bf 8})
\end{array}
\right.
\eeq
entry $\textsl{3}$ corresponds to the $\cN=4$ theory with $\suf(3)$ Lie-algebra\footnote{It is well known that, for a given gauge algebra, there are multiple $\cN=4$ theories \cite{Aharony:2013hda}. Here we are using the determinative article because we are neglecting all the subtleties connected with the global structure of the gauge group.}, while entry \textsl{1} and \textsl{2} correspond to genuinely $\cN=2$ theories. These are all rank-2 theories and we will indicate the two complex coordinates which parametrize their CBs as $(u,v)$.

To derive the discriminant locus we can take advantage of the fact that the theory is lagrangian and directly study the values of $(u,v)$ for which extra massless states enter the theory. To do that it is helpful to write the CB coordinates explicitly in terms of the eigenvalues of $\Phi$, the scalar component of the $\suf(3)$ adjoint $\cN=2$ vector multiplet. First imposing $F$ and $D$ term conditions, constraints $\Phi$ to the be gauge equivalent to the following form:
\beq\label{SolSU3}
\Phi=\left(
\begin{array}{ccc}
a_1&0&0\\
0&a_2&0\\
0&0&-a_1-a_2
\end{array}
\right)
\eeq
and for $a_1\neq a_2$, $\suf(3)\to \U(1)^2$. A convenient choice for $u$ and $v$ is, explicitly:
\beq\label{vevs}
u:={\rm Tr}\big[ \Phi^2\big]\sim a_1^2+a_2^2+a_1a_2,\qquad v:={\rm Tr}\big[ \Phi^3\big]\sim-a_1^2 a_2-a_1a_2^2
\eeq
where we have neglected inessential numerical coefficients. From \eqref{vevs}, it follows straightforwardly that the coordinates of the CB of all three theories in \eqref{theories} have scaling dimensions $\D_u=2$ and $\D_v=3$. 

When two of the eigenvalues in \eqref{SolSU3} coincide, one of the two low energy $\U(1)\to \suf(2)$ and the $\cN=2$ vector multiplets of the corresponding gauge bosons become massless giving rise to singularities on the CB. It is easy for the reader to check that $a_1=a_2$ corresponds to the following hypersurface in $\cC$:
\beq\label{singN2V}
u^3+\l v^2=0,
\eeq
where $\l\in \C^*$ depends on the specific normalization of $u$ and $v$. 

Due to superpotential terms, the hypermultiplets are massive on a generic point of $\cC$ but extra components of the singular locus might arise if special values of $a_1$ and $a_2$ exist for which some of the hypermultiplets components become massless. Here the result depends on the specific representations in \eqref{theories}. Straightforward representation theory shows that components of the hypermultiplets become massless at:
\begin{align}
\left.
\begin{array}{c}
\onesl.\\
\twosl.
\end{array}\right\}&
\quad a_1/a_2=0\quad \Rightarrow\quad v=0\label{a1z}
\\
\threesl.\quad&\quad a_1=a_2\quad\quad \Rightarrow\quad u^3+\l v^2=0.
\end{align}
The fact that components of the hypermultiplets for theory $\threesl$ become massless at \eqref{singN2V} follows from the fact that for this theory the hypermultiplets transform in the same representation as the $\cN=2$ vector multiplets. 

To improve on this semi-classical analysis and write down the discriminant locus, we notice that for theories $\onesl.$ and $\twosl.$ no hypermultiplets become massless at \eqref{singN2V}. The low energy effective theory there is then a $\cN=2$ $\suf(2)$ Yang-Mills and the well-known results of \cite{Seiberg:1994rs} imply that \eqref{singN2V} has to ``split'' in two components accounting for the monopole and dyon singularities.\footnote{The closing of the singular locus under the $\C^*$ action implies that the separation between these two components increases as we move towards the asymptotics of the CB. This is consistent with a careful RG analysis and the original result \cite{Seiberg:1994rs}.} No splitting instead happens for theory $\threesl.$ because the extra massless hypermultiplets component make the low-energy theory on \eqref{singN2V} superconformal as we will see explicitly below.

Bringing all together we find:
\begin{subequations}
\begin{align}\label{DisLo1}
\left.
\begin{array}{c}
\onesl.\\
\twosl.
\end{array}\right\}&
\qquad D_x\sim v (u^3+\l_1 v^2)(u^3+\l_2 v^2)
\\\label{DisLo2}
\threesl.\quad&\qquad D_x\sim (u^3+\l v^2)
\end{align}
\end{subequations}
and therefore we have the following for theories $\onesl.$ and $\twosl.$:
\begin{align}\nonumber
\cSb_1\quad:&\quad\Big\{(u,v)\in\cC\Big| v=0\Big\}\qquad\quad\ \  \qquad\Rightarrow \quad\D_1^{\rm sing}=3,\\\label{sing3}
\cSb_2\quad:&\quad \Big\{(u,v)\in\cC\Big|u^3+\l_1 v^2=0\Big\}\qquad\Rightarrow\quad \D_2^{\rm sing}=6,\\\nonumber
\cSb_3\quad:&\quad \Big\{(u,v)\in\cC\Big|u^3+\l_2 v^2=0\Big\}\qquad\Rightarrow\quad \D_3^{\rm sing}=6.
\end{align}
while theory $\threesl.$ only has a single component analogous to either $\cSb_2$ or $\cSb_3$. We will come back to these examples throughout the manuscript to elucidate the definitions that we make along the way.

\subsection{Low-energy effective theory on the singularity}\label{sec:lowene}

%. Therefore along a given hypersurface $\cSb_i$, the low-energy effective theory is no longer a boring $\U(1)^r$ but rather a conformal or IR-free rank-1 theory whose charged massless states span the sublattice $\L_i$ (times a boring $\U(1)^{r-1}$ factor). The fact that the non-trivial factor is rank-1 is a consequence of the fact that the $\cSb_i$ are complex co-dimension 1

Let's see how we can naturally associate a rank-1 theory to each $i\in I$ indexing the singular components in \eqref{stra1}. This will provide the missing information which we need to use \eqref{actotaint}-\eqref{actotcint} to compute the central charges of any $\cN=2$ SCFT.  

Call the rank-$r$ theory at the superconformal vacuum $\cT$ and call $\cT_\bu$ the low-energy effective description of $\cT$  at the generic point $\bu$ of the CB $\cC$ describing the massless degrees of freedom there. For example we have:
\beq
\cT_\bu\equiv \text{free $\cN=2$ $\U(1)^r$},\qquad \bu\in \cCrg.
\eeq
From what we just discussed, $\cTu$ will include charged states if $\bu\in\cSb$. Each separate component $\cSb_i$ is identified by a lattice of charges $\L_i$ which identifies the charges of these states (those becoming massless there and for which $Z_Q(\bu)\big|_{\cSb_i}$ vanishes). To be precise, a given sublattice is associated to an open subset $\cS_i\subset \cSb_i$ and we will call $\cS_i$ the stratum associated to $\cSb_i$ (which is indeed the closure of $\cS_i$). It is then natural to associate to each stratum the theory describing the massless degrees of freedom on the stratum itself. We will call this theory $\cT_i$ and since $\cS_i$ is complex co-dimension one, it can be shown that $\cT_i$ has to be rank-1. See appendix \ref{app:CB} for clarification on this and other subtleties, like the difference between $\cS_i$ and $\cSb_i$. In summary:
\beq\label{pairT}
\cT_i\equiv \cT_\bu,\qquad \bu\in\cS_i, \quad i\in I.
\eeq
and the quantities indexed by $i\in I$, $(c_i,k_i,h_i)$, which enter the central charge formulae \eqref{actotaint}-\eqref{actotcint} refer to those of $\cT_i$. We also use $u_i$ to label the coordinate parametrizing the one complex dimensional CB $\cC_i$ of $\cT_i$ and define:
\beq\label{Di}
\D_i:=\D(u_i)
\eeq
which defines the last quantity entering the central charge formulae.

\begin{center}
\rule[1mm]{2cm}{.4pt}\hspace{1cm}$\circ$\hspace{1cm} \rule[1mm]{2cm}{.4pt}
\end{center}

\paragraph{Example: SCFTs with $\suf(3)$ gauge algebra} To consolidate \eqref{pairT}, let's see explicitly how this works in the examples \eqref{theories}. We have done most of the work above already. In this rank-2 case, the strata associated to the various components are straightforwardly obtained by subtraction of the origin of $\cC$:
\beq
\cS_i:=\cSb_i\setminus \{0\}.
\eeq
We will clarify momentarily why the rank-1 theories are properly defined on $\cS_i$ and not $\cSb_i$.

Let's start from $\cS_1$ in \eqref{sing3}. For $\{v=0\}\setminus \{0\}$, see \eqref{a1z}, the unbroken gauge group is $\U(1)^2$ with massless hypermultiplets corresponding to the weights associated to the vanishing eigenvalue. It is possible to choose the low-energy $\U(1)$s in such a way that the component of the massless hypermultiplets are only charged under one of the $\U(1)$ factors. Given the $\suf(3)$ representations of the hypermultiplets, it is a straightforward group theory exercise to compute both the number and the $\U(1)$ charges, $q$, of the massless components and we find:
\beq\label{I6poss}
\cT_1:\quad\left\{
\begin{array}{cl}
\onesl\ \to&\, \U(1)\ w/\ 6\ {\rm hypers\ with}\ q=1\\
\twosl\ \to&\, \U(1)\ w/\ 2\ {\rm hypers\ with}\ q=1\oplus 1\ {\rm hyper\ with}\ q=2 \\ 
\end{array}
\right.\quad \Rightarrow\quad \D_1=1,
\eeq 
since the CB scaling dimension for $\cN=2$ $\U(1)$ gauge theories is one. The remaining components in \eqref{sing3} are associated to the monopole/dyon singularities of the $\cN=2$ $\suf(2)$ super Yang-Mills for both theory $\onesl.$ and $\twosl.$. We can separately choose an electromagnetic duality basis for $\cS_2$ and $\cS_3$ which makes the massless degrees of freedom there electrically charged:
\beq\label{th23}
\cT_{2,3}:\qquad \to \quad \U(1)\ w/\ 1\ {\rm hypers\ with}\ q=1\quad \Rightarrow\quad \D_{2,3}=1,
\eeq
This concludes the analysis of theories $\onesl.$ and $\twosl.$. 

For the $\cN=4$ theory we only have to determine the low energy theory on the one locus where one of the low-energy $\U(1)$ factors enhances $\U(1)\to\suf(2)$. Since the hypermultiplet of this theory transforms in the same representation of the $\cN=2$ vector multiplet, we expect the massless states coming from the hyper to fill an adjoint representation of $\suf(2)$ giving in the low-energy limit a rank-1 $\cN=4$ theory:
\beq\label{thN4}
\cT_{\cN=4}:\quad\to\quad \cN=4\ \suf(2)\quad \Rightarrow\quad \D_{\cN=4}=2.
\eeq
The last equation follows from the fact that the CB parameter of an $\suf(2)$ theory is identified with $u\sim \langle{\rm Tr}\Phi^2\rangle$ and therefore $\D=2$.

Now that we have identified the rank-1 theories supported on the various complex co-dimension strata $\cS_i$, we can use the information \eqref{I6poss}-\eqref{thN4} and compute the quantities which enter the central charge formulae for the $\suf(3)$ theories in \eqref{theories} and which will be used in the next section:
\beq\label{cis}
\begin{array}{cc|ccccc}
\multicolumn{2}{c|}{\quad \text{Theory}\quad\,}&\ 12c_i\ &\ \D_i\ &\ \ff_{k_\ff}\ &\ h_i\ &\ T(\boldsymbol{ 2h})\\
\hline
\hline
\multirow{2}{*}{$\cT_1\ \Big\{$}&\onesl.&8&1&\suf(6)_2&0&0\\
\cline{2-7}
&\twosl.&5&1&\varnothing&0&0\\
\hline
\multicolumn{2}{c|}{\cT_{2,3}}&3&1&\varnothing&0&0\\
\hline
\multicolumn{2}{c|}{\cT_{\cN=4}}&9&2&\spf(1)_3&1&1\\
\hline
\hline
\end{array}
\eeq
Here $\spf(1)\cong\suf(2)$ and we have only reported the simple flavor factors of the flavor symmetries. The formula for computing $12c$ for lagrangian theories is standard but it is also recalled below, see \eqref{normack}. Finally $h_i$ here is the quaternionic dimension of the ECB of the theory, that is the number of free hypermultiplets at the generic point of the theory's CB.

Let's conclude with a clarification of why rank-1 theories are associated to the $\cSb_i\setminus\{0\}$. Setting both $u=v=0$ fully restores the $\suf(3)$ gauge algebra. The superconformal vacuum is then associated to a rank-2 theory (the origin is also the only complex co-dimension higher than 1 stratum in this case, see appendix \ref{app:CB}). Therefore a non-trivial rank-1 factor is only defined  away from $\{0\}$, that is strictly on $\cS_i$.

\begin{center}
\rule[1mm]{2cm}{.4pt}\hspace{1cm}$\circ$\hspace{1cm} \rule[1mm]{2cm}{.4pt}
\end{center}

\subsection{Quantum discriminant}\label{sec:QuaDis}

The discriminant locus \eqref{DisLoc} provides a nice algebraic description of $\cSb$ but tells no information about the low-energy theory $\cT_i$ supported on its strata, as we saw in the various examples. If a rank $r$ $\cN=2$ SCFT admits a formulation of special Kahler geometry of its CB in terms of a family of hyperelliptic curves, the Seiberg-Witten (SW) curve $\S$, and one forms, the SW one-form $\l_{\rm SW}$, we can improve on the discriminant locus and define \emph{the quantum discriminant}.  Even though we will not review here this formulation of SW theory,\footnote{Good references are the original papers \cite{Seiberg:1994rs,Seiberg:1994aj} or some of the reviews of SW geometry e.g. \cite{AlvarezGaume:1996mv,Lerche:1996xu,Martone:2020hvy}.} we will still require some technical definitions to arrive at the formulation of the quantum discriminant. The content of this subsection is important to understand the subtleties of how the geometry of the CB can be leveraged to understand the singular locus but will not be needed to understand the rest of the paper.

Bring the curve $\S$ in the following form:
\beq\label{hyper}
\S\quad:\quad y^2=\Psc(x,\bu)
\eeq
where $\Psc(x,\bu)$ is a polynomial of degree, at most, $2r+2$ in $x$ with coefficients which in general depend on $\bu$. Therefore for fixed value $\bu\in\cC$, \eqref{hyper} defines a hyper-elliptic curve $\S_{\bu}$ which degenerates if some of the $2r+2$ roots of $\Psc(x,\bu)$ coincide. The loci $\bu$ for which $\S_\bu$ degenerates corresponds to the singular locus $\cSb$ and is therefore reasonable to define \emph{the quantum discriminant} as the $x$ discriminant of the RHS of \eqref{hyper}:
\beq\label{QuaDis}
D_x^\L:={\rm Disc}_x \Psc(x,\bu).
\eeq
The zeros of the quantum discriminant should coincide with the discriminant locus but \eqref{QuaDis} will be in general no longer reduced:
\beq\label{QuaDisell}
D_x^\L\sim \prod_{i\in I} \Big[P_i(\bu)\Big]^{\ell_i},\qquad \cSb_i=:\Big\{\bu\in\cC\Big| P_i(\bu)=0\Big\}
\eeq 
The $\ell_i\in\Z$ are then the order of vanishing of $D_x^\L$ near the zeros of $P_i(\bu)$.

\begin{table}[t!]
\centering
$\begin{array}{|c|l|c|c|}
\hline
\multicolumn{4}{|c|}{\text{\bf Scaling behaviors near rank 1 singularities}}\\
\hline\hline
\text{Name} & \multicolumn{1}{c|}{\text{planar SW curve}} & \ \text{ord}_0(D^\L_{x})\ \ &\ \D(u)\ \ \\
\hline
II^*   &\parbox[b][0.45cm]{4cm}{$\ y^2=x^3+u^5$}             
&10 &6 \\
III^*  &\ y^2=x^3+u^3x &9 &4 \\
IV^*  &\ y^2=x^3+u^4 &8 &3 \\
I_0^* &\ y^2=\prod_{i=1}^3\left(x-e_i(\t)\, u\right)
&6 &2 \\
IV &\ y^2=x^3+u^2 &4 &3/2 \\
III &\ y^2=x^3+u x &3 &4/3 \\
II  &\ y^2=x^3+u &2 &6/5 \\
\hline
\hline
I^*_n\ \ (n{>}0) &
\parbox[b][0.45cm]{5cm}{
$\ y^2=x^3+ux^2+\L^{-2n}u^{n+3}\ \ $}
& n+6 & 2\\
I_n\ \ (n{>}0)    &\ y^2=(x-1)(x^2+\L^{-n}u^n)  
& n     & 1 \\[0.5mm]
\hline
\end{array}$
\caption{\label{tab:Kodaira} Scaling forms of rank 1 planar special Kahler singularities, labeled by their Kodaira type (column 1), a representative family of elliptic curves with singularity at $u=0$ (column 2), order of vanishing of the discriminant of the curve at $u=0$ (column 3), mass dimension of $u$ (column 4), a representative of the $SL(2,\Z)$ conjugacy class of the monodromy around $u=0$ (column 5), the deficit angle of the associated conical geometry (column 6), and the value of the low energy $\U(1)$ coupling at the singularity (column 7).  The first seven rows are scale invariant.  The last two rows give infinite series of singularities which have a further dimensionful parameter $\L$ so are not scale invariant; they can be interpreted as IR free theories since $\t_0=i\infty$.}
\end{table}

In a rank-1 theory the curve $\S$ is elliptic and it can always be brought in its Weierstrass form before taking its $x$ discriminant. This is how we define the quantum discriminant of a rank-1 theory. Since the singular locus in this case is just the origin, the quantum discriminant will be proportional to $u^\ell$, $u$ being the global coordinate describing the CB of the rank-1 SCFT. It is known that in this case the order of vanishing $\ell$ strongly constrains the scaling dimension $\D(u)$ as can be seen looking at the third column of table \ref{tab:Kodaira}. For instance, if $\ell=1$ then the CB geometry has to be an $I_1$ and $\D(u)=1$. If $\ell=10$ there are instead three possibilities $II^*$, $I_4^*$ or $I_{10}$, which correspond to $\D(u)=6,2,1$, respectively. There is a natural generalization of this fact for the quantum discriminant of higher rank theories and the $\ell_i$ in \eqref{QuaDisell} can be identified with the order of vanishing of the quantum discriminant of the rank-1 theory $\cT_i$ supported on $\cS_i$ and therefore strongly constrains the $\D_i$ in \eqref{Di}.

\begin{center}
\rule[1mm]{2cm}{.4pt}\hspace{1cm}$\circ$\hspace{1cm} \rule[1mm]{2cm}{.4pt}
\end{center}

\paragraph{Example: SCFTs with $\suf(3)$ gauge algebra}  Let's go back to our $\suf(3)$ theories and consider first theories $\onesl$ and $\twosl$. The SW curve for the two theories in terms of a two-parameter family of hyperelliptic curve was derived in \cite{Argyres:1995wt,Landsteiner:1997vd}. When the masses are turned off the two curves coincide:
\beq\label{curSU3}
y^2= x^6+(v+ux)^2+x^3(v+u x)\tau
\eeq
where $\tau$ is the $\suf(3)$ holomorphic gauge coupling. Taking the $x$ discriminant of the RHS of \eqref{curSU3} we can then readily compute its quantum discriminant:
\beq\label{QuaDis2}
\left.
\begin{array}{c}
\onesl.\\
\twosl.
\end{array}
\right\}
\quad:\quad D_x^\L\sim v^6(u^3+\l_1 v^2)(u^3+\l_2 v^2)
\eeq
where $\l_i\in\C^*$ and we left out inessential numerical coefficients. As expected, the zeros of the quantum discriminant coincide with the discriminant locus, see \eqref{DisLo1}, but now we can extract extra information from the order of such zeros:
\beq\label{ordsing}
\ell_1=6,\quad \ell_{2,3}=1.
\eeq
Before interpreting these numbers it is useful to remind the physical interpretation of the $I_n$ and $I_n^*$ geometry in table \ref{tab:Kodaira} which are associated, respectively, to the CB of a $\U(1)$ and a $\suf(2)$ $\cN=2$ gauge theory. The $n$ in both cases refers to the coefficient of the gauge coupling beta function which is:
\begin{align}
\U(1)\quad&:\quad b_g=\sum_iq_i^2\\
\suf(2)\quad&:\quad b_g=\sum_i\Big(T\big({\bf R}^i_{\rm hypers}\big)-4\Big)
\end{align}
where the $q_i$ are the electric charges of the hypermultiplets and $T\big({\bf R}^i_{\rm hypers}\big)$ is the quadratic index of the $\suf(2)$ representation of the $i$-th hypermultiplet.\footnote{We choose a normalization for which $T({\bf 3})=4$.} Using \eqref{I6poss} and \eqref{th23} we can straightforwardly apply these formula to conclude that for both theory $\onesl.$ and $\twosl.$ we expect the CB of $\cT_1$ to be an $I_6$ while that of $\cT_{2,3}$ to be an $I_1$. We leave it as an exercise for the reader to check the matching of the values in \eqref{ordsing} from the corresponding entry in table \ref{tab:Kodaira}.

For theory $\threesl.$ no SW curve is known in the form that we need for the computation of the quantum discriminant. Yet above we have determined the theory supported on the singular locus and could use this information to infer the expected quantum discriminant:
\beq
\threesl.\qquad D_x^\L\sim (u^3+\l v^2)^6
\eeq
where we used \eqref{thN4} and the fact that $T({\bf 3})=4$ and therefore the CB of the theory supported on the singular locus is a $I_0^*$.

\begin{center}
\rule[1mm]{2cm}{.4pt}\hspace{1cm}$\circ$\hspace{1cm} \rule[1mm]{2cm}{.4pt}
\end{center}

As we saw from the examples, the quantum discriminant improves on the discriminant locus and provides considerable extra information about the low-energy limit of the superconformal theory. A big shortcoming of this nice story is that the value of the $\ell_i$'s, which should be an invariant of the special Kahler geometry, is instead not invariant under coordinate reparametrization of the curve \eqref{hyper}. In defining \eqref{QuaDis} we have secretly dealt with this ambiguity by assuming that the curve in \eqref{hyper} is in some sort of higher rank generalization of the Weierstrass form. This is obtained by extending the $\C^*$ action on $\cC$ to $\S_\bu\cong \Csc_r$, fixing the $[x]$ and $[y]$ in terms of $[\bu]$ via the definition of the SW differential:
\beq\label{onefor}
\frac{\del \l_{\rm SW}}{\del u_i}=\frac{x^{i-1} dx}{y^2}:=\w_i,\qquad \bu=(u_1,...,u_r)
\eeq
where the $\w_i$, $i=1,...,r$ is a basis of $H^0(\Csc_r,\Omega^1_{\Csc_r})$, the space of globally defined holomorphic one-forms on $\S_\bu$. Once we appropriately extend the $\C^*$ action, we can harvest the power of scale invariance. It in fact follows that $\Psc(x,\bu)$ in \eqref{hyper} must be a weighted homogeneous polynomial of weight:
\beq
\big[\Psc(x,\bu)\big]=2[y]
\eeq
which immediately fixes the degree $d$ of the highest term in $x$. We can fix the remaining reparametrization invariance setting to zero the term of degree $d-1$ in $x$ though the quantum discriminant does not depend on this last step.

The quantum discriminant is certainly a useful object to consider, yet our definition seems too convoluted for an object whose physical interpretation is instead sharp and appears to be intrinsic to the special Kahler geometry of $\cC$. Furthermore hyper-elliptic curves are a set of measure zero in the space of complex curves of genus $g>2$. Therefore the special Kahler geometry of a generic $\cN=2$ SCFT with rank $\geq 3$ will not have a formulation in these terms and we don't know how to replace the definition of the quantum discriminant in a useful way.  Finding an intrinsic algebraic definition of the quantum discriminant which does not make use of the explicit realization of special Kahler geometry in terms of a family of hyperelliptic curves will bring significant progress in the study of scale invariant CBs of arbitrary ranks. Most importantly, the quantum discriminant is not the natural quantity which enters the twisted partition function.

\subsection{Physical discriminant}\label{sec:PhyDis}

In this subsection we introduce an important definition for the discussion of the next section where we analyze the partition function of topologically twisted $\cN=2$ SCFTs at arbitrary rank. The partition function of a topologically twisted $\cN=2$ SCFT reduces to an integral over the CB parameter $\bu$. The non trivial dependence of the integrand can be fixed by carefully analyzing its dependence near the singular locus $\cSb$.  The \emph{physical discriminant} arises precisely in this context. 

As we will review below, the $\cB$ and $\cC_a$ factors of the measure of the integrand of the partition can only vanish or diverge at the zeros of \eqref{DisLoc}. In particular by the general consideration of \cite{Witten:1995gf,Moore:1997pc,Losev:1997tp,Marino:1997gj}, the $\cB$ factor should be proportional to a close cousin of the quantum discriminant which was called \emph{physical discriminant} in \cite{Shapere:2008zf}:
\begin{align}\label{Bfact}
\cB=\beta \left(D^{\rm phys}_x\right)^{1/8}
\end{align}
where $\b$ is a constant. $D^{\rm phys}_x$ is a close cousin of \eqref{QuaDis} in the sense that we expect $D_x^{\rm phys}$ to share the same zeros as the quantum discriminant and to also be non-reduced. We will parametrize our ignorance in the following way:
\beq\label{DisPhys}
\Dph\sim \prod_{i\in I} \Big[P_i(\bu)\Big]^{b_i},\qquad \cSb_i=:\Big\{\bu\in\cC\Big| P_i(\bu)=0\Big\}
\eeq
which is indeed very resembling of \eqref{QuaDis}]. Determining the form of the $b_i$s will be one of the goals of the next section. It will be the determination of these exponents that leads to our central charge formulae \eqref{actotaint}-\eqref{actotcint}.

\section{Twisted partition function and central charge}\label{sec:Twisted}

$\cN=2$ supersymmetric gauge theories are related via a well-known twisting procedure to topological field theories \cite{Witten:1988ze}, see \cite{Labastida:2005zz,Moore:2017} for a pedagogical presentation of this subject. The starting point is the Euclidean version of $\cN=2$ supersymmetric theories with total symmetry group:
\beq
\cG=\cK\times SU(2)_R\times \U(1)_R
\eeq
where the $\cK=SU(2)_+\times SU(2)_-$ is the rotation group and $SU(2)_R\times \U(1)_R$ the R-symmetry. Under this symmetry the supercharges transform as:
\begin{subequations}
\begin{align}
Q&:\quad ({\bf 1,2,2})^{-1}\\
\tilde Q&:\quad ({\bf 2,1,2})^{1}
\end{align}
\end{subequations}

The topological twisting procedure consists in modifying the coupling to gravity by redefining the spin of the fields. To do this we introduce an external $SU(2)_R$ gauge potential setting it equal to the self-dual part of the spin connection. Or, in other words, re-define the rotation group as $\cK'=SU(2)'_+\times SU(2)_-$ where $SU(2)'_+=\big[SU(2)_+\times SU(2)_R\big]_{\rm diag}$ is the diagonal combination of the two remaining $SU(2)$'s. It is easy to see that under $\cK'\times \U(1)_R$ the supercharges transform:
\begin{align}
Q&:\quad\quad ({\bf 2,2})^{-1}\\
\tilde Q&:\quad ({\bf 3\oplus 1,1})^1
\end{align}
therefore there is one supercharge which is a singlet under the rotation group. We call it $Q_{\rm BRST}$ and it is with respect to this supercharge that the supersymmetry preserved on a curve background (see below) is defined. The topological sector of the twisted theory is then obtained by projecting into the cohomology of the singlet supercharge. The resulting theory has many remarkable properties including the fact that its partition function, discussed in more detail momentarily, does not depend on the metric \cite{Witten:1988ze}. This is the sense in which the theory is topologically invariant.

Upon flowing to the IR on the CB, the partition function of the twisted theory is given by the path integral of the low-energy lagrangian after integrating out massive degrees of freedom. Since the twisted theory is topological, the contributions of the massive states can only result in terms which are proportional to topological densities \cite{Witten:1995gf, Shapere:2008zf}:
\beq\label{Lir}
\cL_{\rm twisted}\sim \cL_\text{IR[V,H]}+\big({\rm log}\,\cA\big){\rm tr} R\wedge\tilde{R}+\big({\rm log}\,\cB\big){\rm tr} R\wedge R+\sum_a\big({\rm log}\,\cC_a\big){\rm tr} F_{\ff_a}\wedge F_{\ff_a}
\eeq
$R$ is the Riemann tensor of the four manifold and $F_{\ff_a}$ is an external gauge field coupled to the simple factor $\ff_a$ of the twisted theory flavor algebra $\ff=\oplus_a\ff_a$. $\cL_{\rm IR[V,H]}$ is the sum of a BRST exact term and a contribution which depends on the coupling of the low-energy Abelian theory on a generic point of the CB. The details of $\cL_{\rm IR[V,H]}$ will not play any role below so we will not write its dependence explicitly and refer to the literature for details \cite{Witten:1995gf,Moore:1997pc}. The factors $(\cA,\cB,\cC_a)$ depend holomorphically on the globally defined CB coordinates $\bu$ and we will spend a considerable part of this section to determine their detailed form. We will henceforth not discuss possible $\U(1)$ flavor factor as in that case equation \eqref{Lir} becomes considerable more subtle and non-holomorphic terms might arise.\footnote{We thank G. Moore for clarifying this point and sharing some unpublished results \cite{Moore:2020}.}

Bringing all together we have the following expression for the twisted partition function \cite{Witten:1995gf, Shapere:2008zf}
\begin{align}\label{Ztwist}
Z=\int [dV] [dH]\ \cA^\ch\ \cB^\s \ \textstyle{\prod_a} \cC_a^{n_a} 
\ e^{S_\text{lR}[V,H]} .
\end{align}
where $\s$ and $\chi$ are the signature and Euler characteristics of the four manifold and $n_a$ is the instanton number for the $\ff_a$ factor. The path integral is over the $n_V$ (IR free) massless neutral vector multiplet fields and $n_H$ massless neutral hypermultiplet fields (if any) on the generic point of the CB $\cC$. When there are free hypers on the generic point of $\cC$, by giving a vev to them, it is possible to see $\cC$ as a part of a larger branch of the moduli space usually called the extended CB (ECB). $n_H$ is then naturally interpreted as the quaternionic dimension of the ECB $h$. $n_V$ is instead naturally identified with the rank $r$ of the theories. Therefore we have:
\beq\label{ECB}
n_H=h,\qquad n_V=r.
\eeq
The integral \eqref{Ztwist} includes an ordinary integral over the 0-modes (constant modes) of the vector multiplet scalars, $u$.

The expression for the $\cA$ factor was determined in full generality to be \cite{Moore:1997pc,Losev:1997tp,Marino:1998bm}:
\begin{align}\label{Afact}
\cA = \a \left[\det\left(\frac{\del u^i}{\del a^j}\right)\right]^{1/2}
\end{align}
where $\a$ is a constant and for a theory of rank-$r$ the indices $i,j$ run up to $r$. The expression for the $\cB$ and $\cC_a$ factor are instead more involved. They can only vanish along the singular locus of the CB and therefore both $\cB$ and $\cC_a$ are somewhat related to the discriminant locus \eqref{DisLoc}. In order to clarify this dependence we need to take a detour and remind the reader about the relationship between the $\U(1)_R$ anomaly of the topologically twisted $\cN=2$ supersymmetric theory and the value of the $(a,c,k_\ff)$ of the untwisted version. This relationship will also allow us to finally derive our central charge formulae. % write, once we determine the form of $\cB$ and $\cC_a$, the $(a,c,k_\ff)$ of any (untwisted) $\cN=2$ SCFT in terms of low-energy data on its CB, see \eqref{actota}, \eqref{actotb} and \eqref{actotc} below.

% and exploiting the fact that the partition function allows to keep track of the $\U(1)_R$ anomaly, which we will label as  as we are going to review now. 

\subsection{$\U(1)$ anomaly and central charges}\label{Uano}

The $a$ and $c$ central charges of the 4d conformal algebra are certain coefficients in OPEs of energy-momentum tensors, and the $k$ central charges appear in the OPEs of flavor currents. We will assume throughout that the flavor algebra $\ff = \oplus_a \ff_a$ is a sum of simple factors and each factor will have a separate $k_a$ central charge. The fact that these quantities appear in the $\D R$ anomaly follows from the fact that $(a,c,k_a)$ appear in the scale anomaly in the presence of a background metric and background gauge fields for $\ff$. Then $\cN=2$ superconformal symmetry relates the scale anomaly to 't Hooft anomalies for the $\U(1)_R\oplus\SU(2)_R\oplus_a\ff_a$ global symmetry, with the result that in the presence of a background metric and background gauge fields for the global symmetries, the conservation of the $\U(1)_R$ current is broken by terms proportional to the central charges times topological densities formed from the background fields. Notice that background metric and gauge fields describing an arbitrary smooth oriented 4-fold $M$ with $F$-bundle (where $F$ is the flavor symmetry group with Lie algebra $\ff$) generally break $\cN=2$ supersymmetry.   However, as we reviewed above, the twisted topological version of it is still protected by a supersymmetry \cite{Witten:1988ze}, despite a curve background, and therefore we can still talk about a $\U(1)_R$. The result \cite{Witten:1995gf,Shapere:2008zf} is that the partition function of the twisted theory on $M$ with an $F$-bundle carries $\U(1)_R$ charge\footnote{Here we are using a normalization of the $\U(1)_R$ charge such that $R(u) = \D(u)$.  This differs from that used in \cite{Shapere:2008zf} by a factor of two.}
\begin{align}\label{DRxsn}
\D R = (2a-c)\cdot \chi + \frac32 c \cdot \s - \frac12 \sum_a k_a \cdot n_a,
\end{align}
where $\chi$ and $\s$ are the Euler characteristics and signature of $M$ and $n_a$ are the instanton numbers of the $F$-bundle. Here we have again assumed that the flavor symmetry of the SCFT is a semi-simple Lie algebra $\ff = \oplus_a \ff_a$ and $k_a$ is the level of flavor currents of the simple factor $\ff_a$.

\eqref{DRxsn} corresponds to the standard normalizations of the central charges where for $n_V$ free vector multiplets and $n_H$ free hypermultiplets
\begin{align}\label{normack}
24 a &= 5 n_V + n_H, &
12 c &= 2 n_V + n_H, &
k_a &= T_a({\bf2}\bh).
\end{align}
Thus, in this case
\begin{align}\label{freeDRxsn}
\D R_\text{free} = \frac14 n_V \cdot \chi 
+ \left(\frac14 n_V + \frac18 n_H \right)\cdot\s 
- \frac12 \sum_a T_a({\bf2}\bh) \cdot n_a.
\end{align}
Here ${\bf2}\bh$ is the (reducible) representation of $\ff$ under which the $2n_H$ half-hypermultiplets transform.  $T_a({\bf2}\bh)$ is the quadratic index of ${\bf2}\bh$ with respect to the $\ff_a$ factor.\footnote{If ${\bf2}\bh$ decomposes into irreps of $\oplus_{a=1}^L \ff_a$ according to ${\bf2}\bh = \oplus_\a (\br_{\a1}\otimes\br_{\a2}\otimes\cdots\otimes\br_{\a L})$, then $T_a({\bf2}\bh) = \sum_\a \left(\prod_{b\neq a} r_{\a b} \right) T(\br_{\a a})$. The quadratic index for a simple factor is proportional to the sum of the squared-lengths of weights in ${\bf2}\bh$, $T({\bf2}\bh) := (1/\text{rank}\ff)\sum_\l (\l,\l)$, where the weights are normalized so that the long roots of $\ff$ have length-squared 2.  This is the normalization for which $T(\bn)=1$ for $\SU(n)$.}    In case $n_H=0$, there is no contribution from the last term in \eqref{freeDRxsn}, so we adopt the convention that $T(``{\bf 0}"):=0$.

From the explicit expression of the partition function \eqref{Ztwist} its total $\U(1)_R$ charge can be evaluated at a generic (i.e., non-singular) point on the CB to be
\begin{align}\label{DRABC}
\D R = \left(R(\cA) + \frac14 n_V \right)\cdot \chi 
+ \left(R(\cB) + \frac14 n_V + \frac18 n_H \right)\cdot\s 
+ \sum_a \left(R(\cC_a) - \frac12 T_a({\bf2}\bh) \right)\cdot n_a,
\end{align}
where we have used \eqref{freeDRxsn} to evaluate the contribution from the $[dV][dH]$ measure and have left the $R$-charge of the $(\cA,\cB,\cC_a)$ factors to be determined.  Comparing this to \eqref{DRxsn} for arbitrary $(\ch,\s,n_a)$ and using \eqref{ECB} gives \cite{Shapere:2008zf}:
\begin{subequations}
\begin{align}\label{ackia}
24 a &= 5 r + h + 12 R(\cA) + 8 R(\cB) ,
\\\label{ackib}
12 c &= 2 r + h + 8 R(\cB) ,
\\
k_a &= T_a({\bf2}\bh) - 2 R(\cC_a) .
\label{ackic}
\end{align}
\end{subequations}
Let's stress again that the twisting procedure is a purely calculational tool for us, the $(a,c,k_a)$  above refer to the corresponding quantities of the untwisted, non-topological $\cN=2$ SCFT in flat space. 

Computing $R(\cA)$, from \eqref{Afact}, is a straightforward task. Indeed $a^j$, $j=1,...,r$, are the special coordinates on the CB (see appendix \ref{app:CB}) which have scaling dimension one, and thus, in our conventions, $\U(1)_R$ charge one.  The prefactor $\a$ is $u$-independent.  In the conformal case it can only depend on constants which are all dimensionless, so $R(\a)=0$.  We therefore have
\begin{align}\label{RofA}
R(\cA) = \frac{\sum_{i=1}^r \D_{\bu_i}-r}2 
\end{align}
where $\D_{\bu_i}$ is the scaling dimension of the $i$-th entry of the vector of $r$ globally defined complex coordinates on the CB $\bu$. We turn now to a detailed discussion of the $\cB$ factor.

\subsection{Final form of the partition function and central charge formulae}
%$R(\cA)$, $R(\cB)$ and $R(\cC_a)$ to derive from \eqref{ackia}-\eqref{ackic} general formulae for $(a,c,k_a)$ in terms of CB properties. Since we are going to discu 

We already anticipated, as it has been argued for many years, that the $\cB$ factors has to have the following form:
\begin{align}
\cB=\beta \left(D^{\rm phys}_x\right)^{1/8}
\end{align}
and the form of $D^{\rm phys}_x$ was written explicitly in \eqref{DisPhys}. We will turn now to the computation of the $b_i$. 

When the twisted theory is put on a smooth spin 4-manifold, $\cB^\s$ has to be single-valued functions on the CB.\footnote{For non-spin 4-manifolds the $\cB^\s$ measure factor may be multi-valued on the CB \cite{Witten:1995gf}.} Since for smooth spin 4-manifolds $\s\in16\Z$ \cite{rohlin52,teichner92}, we see that $\cB^{16}$ must be a single-valued holomorphic function of $(u,v)$. This immediately implies that
\beq
b_i\in \frac{\Z}2.
\eeq
Recall that the $\ell_i$ in \eqref{QuaDis} are instead integers. To determine the quantities $b_i$ explicitly, we will zoom in on the various complex co-dimension one strata of the singular locus $\cS_i$. There, we can exploit the fact that the $\U(1)_R$ of $\cT_i$ (the rank-1 theory supported on the singular stratum $\cS_i$) provides an accidental low-energy $\U(1)_R$ symmetry which constraints the integrand of the partition function in \eqref{Ztwist} by its anomaly $\D R$. Since this anomaly depends on the physical properties of the low-energy effective theory $\cT_i$, we will be able to relate the $b_i$ to properties of rank-1 theories. This strategy is very resembling of that followed in \cite{Witten:1995gf}.

To do that, consider probing $\cS_i$ away from the origin. We can choose a set of coordinates $(\bu_\parallel,u_\perp)$ on $\cC$ which are ``adapted'' to $\cS_i$. $u_\perp$ is a coordinate transverse to $\cS_i$ while $\bu_\parallel$ are an $r-1$ dimensional vector of coordinates \emph{along} $\cS_i$ in the following sense:\footnote{Zeros of a polynomial form a closed set, therefore $P_i(\bu)=0$ defines $\cSb_i$ not $\cS_i$. Here we focus on $\cS_i$ because we are interested in the set which supports a well-defined $\cT_i$. Therefore, to be precise and a bit pedantic, we should restrict to those values of $\bu_\parallel$ such that $P_i(\bu_\parallel,u_\perp)=0$ spans $\cS_i$.}
\beq
P_i(\bu_\parallel,u_\perp)=0, \quad {\rm for}\quad \{u_\perp=0,\bu_\parallel\in\C^{r-1}\}.
\eeq
The CB parameter of the rank-1 theory $\cT_i$, which we indicated as $u_i$, is a natural choice, at least locally, for $u_\perp$ and therefore we will pick $u_\perp\equiv u_i$, this point is explained in more detail in appendix \ref{sec:stra}. We can expand \eqref{DisPhys} around $u_i=0$ finding:
\beq\label{icomp}
\Dph\sim u_i^{b_i} \cdot \mP(u_i,\bu_\parallel)
\eeq
where $\mP(\bu_\parallel,u_i)$, as long as $\bu_\parallel\neq0$, does not vanish as $u_i \to 0$.%, that is as we restrict $D_x^{\rm phys}$ to $\cS_i$. 

As it is explained in appendix \ref{sec:U1R}, only the $u_i$ is charged under the $\U(1)_R$ symmetry arising near $\cS_i$ and the $\U(1)_R$ charge of the $u_i$ is determined by its scaling dimension $\D_i$ defined in \eqref{Di}:
\beq\label{iRch}
R(u_i)=\D_i,\qquad R(\bu_\parallel)=0.
\eeq

Bringing \eqref{Bfact}, \eqref{icomp} and \eqref{iRch} together we conclude that\
\beq
\frac{b_i \D_i}8 =R(\cB)\big|_i
\eeq
where $R(\cB)\big|_i$ indicates the condition that the $\cB$ $R$-charge has to satisfy near $\cS_i$. From \eqref{ackib} we can solve for $R(\cB)\big|_i$ and get:
\beq
R(\cB)\big|_i=\frac{12 c_i-2-h_i}8
\eeq
where the $_i$ subscript indicates that the given quantity refers to the rank-1 theory $\cT_i$, see for example the quantities calculated in \eqref{cis}.\footnote{Notice that to obtain the formula above we used $n_V=1$ in plugging the value from \eqref{ackib} since the theory $\cT_i$ is a rank-1 theory.}  We then finally get an expression for $b_i$:
\beq\label{bi}
b_i= \frac{12 c_i-2-h_i}{\D_i}
\eeq
The quantity in \eqref{bi} has already appeared in \cite{Argyres:2016xmc}\footnote{The $b_i$ in \eqref{bi} and those defined in \cite{Argyres:2016xmc} differ by an inessential factor of 2 due to a different definition of the physical discriminant.} and has a clear physical significance of being linked to the deformation pattern of the rank-1 theory $\cT_i$ supported on the particular stratum of the singular locus. In other words the value of $b_i$ is related to the way in which $\cT_i$ behaves under relevant deformations. The most promising consequence of this observation is that the study of the mass deformations of higher rank scale invariant CBs might ultimately reduce to a clever application of techniques which were thoroughly analyzed and understood in the context of rank-1. What we are going to learn next studying $\cC_a$, will further support this picture.

Before turning to $\cC_a$, let us write explicitly $R(\cB)$. From \eqref{Bfact} and apply the analysis above to all co-dimension one components we obtain:
\beq
8 R(\cB)=\sum_{i\in I}\D^{\rm sing}_ib_i=\sum_{i\in I}\D^{\rm sing}_i \frac{12 c_i-2-h}{\D_i}.
\eeq
where $\D^{\rm sing}_i$ is defined in \eqref{Dsing}. Notice that the fact that the \eqref{bi} must be semi-integers, can be converted in non-trivial constraints on the rank-1 theories $\cT_i$ which have been discussed in detail in \cite{Argyres:2016xmc}.

We have now determined $R(\cA)$ and $R(\cB)$ and we can finally use \eqref{ackia}-\eqref{ackib} to derive the first two central charge formulae which relate the $(a,c)$ of a SCFT of arbitrary rank $r$ to known quantities of the rank-1 theories supported on its CB singular locus:
\begin{subequations}
\begin{align}
\label{actota}
24 a &= 5r + h + 6 \Big(\sum_{\ell=1}^r\D_{\bu_\ell}-1\Big) +\sum_{i\in I}\D^{\rm sing}_i \frac{12 c_i-2-h_i}{\D_i} ,
\\\label{actotb}
12 c &= 2r + h + \sum_{i\in I}\D^{\rm sing}_{i} \frac{12 c_i-2-h_i}{\D_i},
\end{align}
\end{subequations}
where $i\in I$ labels the complex co-dimension one strata of $\cC$. We will discuss the flavor level, and in general the flavor structure of $\cN=2$ SCFTs in the next section.

\begin{center}
\rule[1mm]{2cm}{.4pt}\hspace{1cm}$\circ$\hspace{1cm} \rule[1mm]{2cm}{.4pt}
\end{center}

\paragraph{Example: SCFTs with $\suf(3)$ gauge algebra}  Let's apply the formulae that we just derived to our $\suf(3)$ examples. In particular we will see that while theories $\onesl.$ and $\twosl.$ shared the same quantum discriminant they will have a different physical discriminant. The calculation here is for pedagogical purposes only since we won't learn anything new on these very well studied lagrangian theories. We will leverage the power of our central charge formulae in \cite{Argyres:2020wmq} where we compute many new properties of $\cN=2$ SCFTs of rank-2 theories and their moduli space. 

Let's start reminding the reader of the properties of these theories:
\beq\label{summ}
\begin{array}{c|ccc}
&\ 24a\, \, &\ 12c\, \,&\ \ff_{k_\ff}\, \, \\
\hline
\hline
N_f=6&58&34&\suf(6)_6\\
1({\bf 6})\oplus 1 ({\bf 3})&49&25&\varnothing\\
\cN=4&48&24&\spf(1)_8\\
\end{array}
\eeq
These can be straightforwardly computed using \eqref{normack} and counting the number of hypermultiplets and vector multiplets. We will reproduce these values from the structure of the CB, in particular using the quantities in \eqref{cis}. It will be a rewarding exercise. Here we will focus on the values for $(a,c)$ while the  flavor part will be discussed in the next section.

Following \eqref{DisPhys}, we can parametrize the physical discriminants of the three theories as:
\begin{align}
\left.
\begin{array}{r}
\onesl.\\
\twosl.
\end{array}
\right\}&\quad:\quad
\Dph\sim v^{b_1^{\onesl,\twosl}}(u^2+\l_1 v^3)^{b_2^{\onesl,\twosl}}(u^2+\l_2 v^3)^{b_3^{\onesl,\twosl}}\\
\threesl.\quad&\quad:\quad
\Dph\sim (u^2+\l v^3)^{b^{\cN=4}}.
\end{align}
We have already determined the low-energy theory along each component of the singularity, see \eqref{I6poss}, \eqref{th23} and \eqref{thN4}, therefore we can directly use \eqref{cis} and equation \eqref{bi} to find the following result:
\begin{align}\nonumber
b_1^\onesl&=6,\quad\quad b_1^\twosl=3,\\\label{bN4}
&b_2^{\onesl,\twosl}=b_3^{\onesl,\twosl}=1,\\\nonumber
&\quad b^{\cN=4}=3.
\end{align}
From \eqref{DisPhys}, this immediately gives the expression for the physical discriminant of three theories:
\begin{align}
\onesl.\qquad \Dph&\sim v^6(u^3+\l_1 v^2)(u^3+\l_2 v^2)\\
\twosl.\qquad\Dph&\sim v^3(u^3+\l_1 v^2)(u^3+\l_2 v^2)\\
\threesl.\qquad\Dph&\sim (u^3+\l v^2)^3
\end{align}
A quick comparison with \eqref{QuaDis2} shows that the quantum and physical discriminants coincide only in one of the three cases.\footnote{There is a clear generalization of this fact, physical and quantum discriminant coincides iff the CB of all rank-1 theories supported on all $\cS_i$'s deform, turning on all relevant deformations, into $I_1$ singularities \cite{Argyres:2015ffa,Argyres:2015gha,Argyres:2016xua,Argyres:2016xmc}.}.

Now let's tackle the central charges of these theories. In this, $\suf(3)$, case, $\D_u$=2 and $\D_v=3$ and $r=2$ so \eqref{actota} and \eqref{actotb} reduce to:
\begin{subequations}
\begin{align}
\label{actota}
24 a &= 10 + h +18 +\sum_{i\in I}\D^{\rm sing}_i \frac{12 c_i-2-h_i}{\D_i}=28 + h +\sum_{i\in I}\D^{\rm sing}_i b_i ,
\\\label{actotb}
12 c &= 4 + h + \sum_{i\in I}\D^{\rm sing}_{i} \frac{12 c_i-2-h_i}{\D_i}=4 + h + \sum_{i\in I}\D^{\rm sing}_{i} b_i,
\end{align}
\end{subequations}
Starting from theory $\threesl.$, the singular locus has only one component (see \eqref{DisLo2}) and the sum will reduce to a single term. This is a rank-2 $\cN=4$ theory, therefore there it has $h=2$:
\begin{align}
24 a_{\cN=4}&= 30+\D^{\rm sing}\, b^{\cN=4}\\\
12 c_{\cN=4}&= 6+\D^{\rm sing}\, b^{\cN=4}
\end{align}
From \eqref{DisLo2}, $\D^{\rm sing}=6$ and we beautifully reproduce the value for $a$ and $c$ in \eqref{summ}. A very similar calculation can be performed for theory $\onesl.$ and $\twosl.$, but now the sum in the central charge formulae includes three factors. We leave it up to the reader to check that things work neatly.

\section{Flavor structure of general $\cN=2$ SCFTs}\label{sec:flavor}

Recall that in a SCFT with flavor symmetry $\ff=\oplus_a\ff_a$, where we restrict  to $\ff_a$ simple, the term $\cC_a$ arises from the contribution to the low-energy lagrangian \eqref{Lir} proportional to the topological density:
\beq\label{flavTop}
{\rm Tr} F_{\ff_a}\wedge F_{\ff_a}
\eeq     
and $F_{\ff_a}$ is an external gauge field coupled to the global symmetry factor $\ff_a$. As discussed above, $\cC_a$ is holomorphic in $\bu$ by topological invariance but, to determine the form of $\cC_a$, we have to understand how the flavor symmetry $\ff_a$ acts on the theory at a generic point of the CB. Notice that all CB operators are uncharged under the flavor symmetry therefore no spontaneous breaking takes place anywhere on $\cC$. As a consequence, the full spectrum at any point of CB, including massive and massless states, should be organized in representations of the full flavor symmetry $\ff$. But each simple factor $\ff_a$ might or might not act on the light states. Determining this is particularly relevant as the zeros of $\cC_a$ can only lie along those components of $\cSb$ where the massless states are charged under $\ff_a$.

This is an important point so let's elaborate on it further. From the definition of the $\cC_a$ factor \eqref{Lir}, the zeros of $\cC_a$ correspond to loci where the contribution to the low-energy Lagrangian, proportional to the topological density \eqref{flavTop}, diverges. This can only happen at the zeros of the discriminant locus \eqref{DisLoc}. But, more specifically, a diverging contribution proportional to the topological quantity \eqref{flavTop} can only be generated if the states becoming massless are charged under $\ff_a$, thus the previous statement. We label the special strata where this happens as $\cS_i$ with $i\in I_{\ff_a}$. Notice that for $i\in I_{\ff_a}$, the rank-1 theory $\cT_i$, describing the massless excitations on $\cS_i$, is strongly constrained by the fact that its flavor symmetry should contain a $\ff_a$ factor.

From these considerations and the fact that all the components of the singular locus can be algebraically parametrized using the polynomial $P_i(\bu)$ in \eqref{DisLoc}, it then follows that $\cC_a$ should have the general form:
\beq\label{Cfact}
\cC_a\sim\prod_{i\in I_{\ff_a}}\big[P_i(\bu)\Big]^{e^a_i},
\eeq
where the $e^a_i$ are for the moment arbitrary. One might wonder whether it is possible that only a subalgebra $\tilde{\ff}_a\subset \ff_a$ is realized on the massless spectrum. But $\ff_a$ is simple and since we argued that it is nowhere spontaneously broken, this possibility is forbidden. An IR flavor symmetry enhancement is instead possible.

Let's now study the zeros of $\cC_a$. First let's indicate the cardinality of $I_{\ff_a}$ as $\nf_a\geq0$. $\cC_a$ is a holomorphic function of $\bu$ and $\cC$ is a connected complex manifold ($\cSbcp=\varnothing$). This implies that the zeros of $\cC_a$ have to be in complex co-dimension one. ${\ff_a}$ acts on the massless spectrum at the superconformal vacuum by assumption, therefore $\cC_a$ vanishes at $\bu=0$. This is enough to conclude that $\cC_a$ has to have a non trivial dependence on at least one of the $P_i(\bu)$ and, for any simple factor $\ff_a\subset\ff$ of the flavor symmetry, $\nf_a\neq0$ . In other words we have just proven the following remarkable fact:\\

\begin{tcolorbox}
\begin{fact}
The Coulomb branch of any $\cN=2$ SCFT with a semi-simple global symmetry $\ff=\oplus_a \ff_a$ has, for each factor $\ff_a$, one or multiple connected co-dimension one strata $\cS_i$ where the charged massless spectrum is organized in non-trivial irreducible representations of  $\ff_a$.
\end{fact}
\end{tcolorbox}\vspace{0.5em}

Since mass deformations of $\cN=2$ theories can be interpreted as background field configurations of the scalar component of a vector multiplet of a weakly gauged complexified flavor algebra $\ff$, the above facts proves the following general condition:\\

\begin{tcolorbox}
\begin{center}
\textbf{UV-IR simple flavor condition}
\end{center}\vspace{-.5em}

All mass deformations of a rank-$r$ $\cN=2$ SCFT deform the CB asymptotically and are realized, in the low-energy limit, as mass deformations of rank-1 theories supported on special complex co-dimension one loci.
\end{tcolorbox}\vspace{.5em}

%\red{When there is a flavor symmetry enhancement it is possible that $\ff_a\subset \ff^{\rm IR}_a$ has a non-trivial index of embedding, therefore we need to account for this. Also the conjecture only applies to non-abelian factors}

We are now in a position to finally complete our discussion on the structure of the partition function. First, using \eqref{ackic} and \eqref{Cfact}, we can immediately obtain an expression for $k_a$:
\beq
k_a=T_a({\bf2}\bh)-2\sum_{i\in I_{\ff_a}} \D_{i}^{\rm sing} e_i^a ,
\eeq
where $\D_{i}^{\rm sing}$ is again defined in \eqref{Dsing} and the sum is restricted to $I_{\ff_a}$. To determine the $e_i^a$ we notice that for \eqref{Ztwist} to be well defined on the CB, $\cC_a^{n_a}$ has to be a single valued function of $\bu$. Since the instanton numbers $n_a\in\Z$ (at least for simply-connected flavor groups), $\cC_a$ itself must be a single-valued holomorphic function of $\bu$ and therefore $e^a_i\in \Z$. 

To determine the value of individual $e_i^a$ we can again zoom-in on the individual complex co-dimension one strata $\cS_i$, as we did for the $b_i$, and apply \eqref{ackic} to the rank-1 theory $\cT_i$ supported there. In doing so, here we restrict to $i\in I_{\ff_a}$ since we are computing the level corresponding to the $\ff_a$ factor. Following precisely the same steps which led us to the determination of the $b_i$s we obtain:
\beq\label{eia}
e_i^a=\frac{T_a({\bf2}\bh_i)-k_a^i}{2d^a_i\D_i}\in \Z,\qquad i\in I_{\ff_a}.
\eeq
Here $k_a^i$ is the level of the $\ff_a$ flavor factor of the rank-1 theory $\cT_i$, $\D_i$ is the scaling dimension of its CB parameter, and $T_a({\bf2}\bh_i)$ is the quadratic index of its ECB. To account for a possible enhancement of the flavor symmetry in the IR, \eqref{eia} also depends on $d_i^a$, the index of embedding of $\ff_a$ into the appropriate simple factor of the flavor symmetry of $\cT_i$.

Putting all together we obtain the last of our central charge formula which determines the level of the higher rank $\cN=2$ SCFT in terms of the levels of the rank-1 theories supported on the complex co-dimension one strata of its singular locus:
\beq
\label{actotc}
k_a=\sum_{i\in I_{\ff_a}}\frac{\D_i^{\rm sing}}{d^a_i\D_i} \left(k_a^i-T_a({\bf2}\bh_i)\right)+T_a({\bf2}\bh).
\eeq
Notice that the fact that the \eqref{eia} must be integers, can be converted into non-trivial constraints which have been discussed in detail in 
\cite{Argyres:2016xmc}.

\begin{center}
\rule[1mm]{2cm}{.4pt}\hspace{1cm}$\circ$\hspace{1cm} \rule[1mm]{2cm}{.4pt}
\end{center}

\paragraph{Example: SCFTs with $\suf(3)$ gauge algebra} As we did throughout the manuscript, let's apply what we just learned to the $\suf(3)$ examples. In particular we will:
\begin{itemize}
\item[$i.$] Check that the $\cN=2$ UV-IR simple flavor condition is indeed respected. 
\item[$ii.$] Compute the flavor level in \eqref{summ} directly from CB quantities.
\end{itemize}

To do the calculation of the flavor level we will analyze theory $\onesl.$, that is $\suf(3)$ with $N_f=6$ which has an $\suf(6)$ flavor symmetry at level 6, see \eqref{summ}. This result can be obtained using \eqref{normack} and observing that the hypermultiplet fileds transform in the fundamental of $\suf(3)$ and both the $\cN=1$ chiral and anti-chiral part contribute a $T({\bf 6})=1$ to \eqref{normack}. From our previous CB analysis, two of three strata support a $\U(1)$ with a single hypermultiplet which carries no-simple flavor factor. The third stratum supports instead a $\U(1)$ gauge theory with six massless hypermultiplets which carries a $\suf(6)$ flavor symmetry, see \eqref{cis}. This checks that the $\cN=2$ UV-IR simple flavor condition is indeed satisfied.

To compute the level of the $\suf(6)$ flavor symmetry of the $\suf(3)$ with $N_f=6$, notice that \eqref{actotc} has an explicit dependence on $\D^{\rm sing}_i$. Therefore to reproduce the proper flavor level we need more information about the stratum where massless states carrying a $\suf(6)$ appear. From our previous analysis we know that the $\U(1)$ theory carrying an $\suf(6)$ is supported on the first entry of \eqref{sing3}, implying $\D^{\rm sing}_1=3$.  Finally, from \eqref{cis}, the level of the $\suf(6)$ of this $\U(1)$ theory is two, $k^1=2$ (the factor of three difference between the level of the $\suf(6)$ in the $\suf(3)$ theory and this case arises because now the hypers are no longer in a three dimensional representation but in a one-dimensional one). It is a matter of elementary algebra to obtain:
\beq
k_{\suf(6)}=\frac{3}1 2=6.
\eeq
Theory $\twosl.$ carries no simple flavor symmetry, but a similar calculation can be carried out successfully for the $\cN=4$ $\suf(3)$.

\section{Conclusions}

In this paper we have thoroughly analyzed the dependence on the singularity structure of the Coulomb branch of the integrand of the partition function for an arbitrary topologically twisted $\cN=2$ superconformal field theory. This exercise allowed us to derive general formulae, \eqref{actota}, \eqref{actotb} and \eqref{actotc}, relating the $(a,c,k_\ff)$ conformal central charges of untwisted $\cN=2$ SCFTs to analogous quantities of rank-1 theories appearing in the low-energy description of the SCFTs at special loci of the their Coulomb branches. These formulae are valid for SCFTs of arbitrary rank and extend and generalize the work of Shapere and Tachikawa \cite{Shapere:2008zf}. The main input which allowed this generalization is our improved understanding of the structure of the stratification of the singular locus of the Coulomb branch \cite{Argyres:2020wmq}.

 In performing our analysis we sharpen the definition of the discriminant of the Coulomb branch and its relevance for understanding the low-energy limit of $\cN=2$ SCFTs. We are also able to derive general lessons about the flavor structure of $\cN=2$ SCFTs and, in particular, on how the simple factors of their flavor symmetry are realized in the low-energy. Our main result in this regard is summarized in the $\cN=2$ \emph{UV-IR simple flavor condition}.

The work described here as well as in a companion paper \cite{Argyres:2020wmq}, is motivated by the classification program of four dimensional $\cN=2$ superconformal field theories \cite{Argyres:2020nrr} based on the systematic analysis of their Coulomb branch geometries. Particularly with the goal of extending the success story of rank-1 \cite{Argyres:2015ffa,Argyres:2015gha,Argyres:2016xua,Argyres:2016xmc} to arbitrary rank. At rank-$r$, this classification program involves two broad steps:
\begin{itemize}
\item[1.] The characterization of all scale invariant CBs of complex dimension $r$.

\item[2.] The analysis of mass deformations of the entries in item $1.$

\end{itemize}

\noindent The result described in this paper represent a, somewhat unexpected, progress on both of these fronts. The $\cN=2$ UV/IR simple flavor condition de-facto reduces the analysis of mass deformations of higher rank theories to a, perhaps involved, implementation of known rank-1 techniques. This therefore identifies a strategy to tackle item $2.$ for arbitrary ranks. Major progress in tackling item $1.$ is represented by the central charge formulae \eqref{actota}, \eqref{actotb} and \eqref{actotc}. Even more so, our detailed understanding of the stratification of the CB singular locus presented in \cite{Argyres:2020wmq} lays the foundation for organizing a search of scale invariant special Kahler geometries of arbitrary complex dimensions. In \cite{Argyres:2020wmq} we will also present a more thorough report on the status of our classification program and the prospects to extend the work beyond rank-1. 

\begin{comment}

Outline simply that this and the companion paper represent a remarkable step forward in our understanding higher rank theories. Indeed by find a way to effectively understand the structure of the stratification of the singular locus of the CB, we are able to relate the analysis of higher rank SCFTs to quantities of rank one theories which have instead be understood in full detail.

Give an update on the status of the classification program, in particular:
\begin{itemize}
\item[1.] Understand scale-invariant geometries: a revisitation of the analysis of \cite{Argyres:2005pp,Argyres:2005wx} using new results will be presented in \cite{Argyres:2020}.

\item[2.] The current work combined with the analysis in \cite{Argyres:2020wmq}, represents a remarkable advancement in the understanding of mass deformations of rank theories. In particular the emergence of the deformation patterns of rank-1 theories to compute the central charges of arbitrary rank theories, suggests that all the mass deformations are indeed realized as deformations of the low-energy rank-1 theories. Since the latter have been fully understood we have reduced the study of a very complicated problem to one which has been already solved.

\end{itemize}

\end{comment}

%!TEX root=../main.tex

\acknowledgments

The author would like to thank Fabio Apruzzi, Philip Argyres, Dnyanesh Kulkarni, Madalena Lemos, Greg Moore, Leonardo Rastelli, Andrea Rocchetto, Sakura Schafer-Nameki and Yuji Tachikawa for helpful conversations, useful suggestions and insightful comments on the draft. The work of the author is supported by NSF grants PHY-1151392 and PHY-1620610.

\appendix

\section{Coulomb branch generalities}\label{app:CB}

Because of the unbroken low energy $\U(1)^r$ gauge invariance on $\cC$, states in the low-energy theory are labeled by a set of $2r$ integral electric and magnetic charges, $Q\in\Z^{2r} = \L$, the \emph{charge lattice}. The $Q$s satisfy the standard Dirac-Zwanziger-Schwinger quantization condition
\beq\label{DirQua}
\langle Q,Q'\rangle\in\Z
\eeq
where the Dirac pairing $\langle \cdot,\cdot\rangle$ gives the charge lattice an integral symplectic structure.\footnote{The situation is slightly more subtle as in general $\langle Q, Q'\rangle:=Q^T\DD Q'$ where $\DD$ is an integer non-degenerate skew-symmetric $2r\times 2r$ matrix which is called the \emph{polarization} of the charge lattice. Here we will always assume it to have its canonical form:
\begin{align}\label{prinpol}
\DD=\bpmat 0 & \I_r\\-\I_r & 0 \epmat 
\end{align}
The choice of $\DD$ in \eqref{prinpol} is also called \emph{principal}. The physical meaning of alternative choices of $\DD$ remains largely unclear. } The matrix of low-energy holomorphic gauge couplings $\t_{ij}$ of the $r$ $\U(1)$ factors at a generic point of $\cC$ can be extracted by writing the Kahler metric on $\cC_{\rm reg}$ in a special basis $ds^2 = \Im ( da^D_j d\bar a^j)=\Im(\t_{ij})da^id\bar{a}^j$, where:
\begin{align}\label{SKconds}
\t_{ij} &:= \frac{\del a^D_i}{\del a^j} = \t_{ji}, &
&\text{and} &
\Im(\t_{ij}) & \ \text{is positive definite.}
\end{align}
$\ba^D$ and $\ba$ are separately good holomorphic coordinates on $\cCrg$ and are the celebrated \emph{special coordinates}:
\begin{align}\label{specCo}
\s &:= \bpmat \ba^D\\ \ba\ \ \epmat,& 
&\text{with}&
\ba^D & := \bpmat a^D_1\\ \vdots\\ a_r^D \epmat, &
&\text{and}&
\ba &:=\bpmat a^1\\ \vdots\\a^r\epmat .
\end{align}

A central fact about CB geometry is that there is no globally defined lagrangian description of the low energy $\cN=2$ $\U(1)^r$ gauge theory, and non-trivial monodromies have to be considered to describe the physics on $\cCrg$.  While the charge lattice remains the same over all points in $\cCrg$, upon dragging a given $Q\in\L$ along a closed path $\g\subset \cCrg$, $Q$ in general suffers a monodromy $Q\overset{\g}{\rightsquigarrow} Q' = MQ $, where $M\in \Sp(2r,\Z)$. $\s$ also suffers monodromies around closed loops and it is therefore a holomorphic section of a rank-$2r$ complex $\Sp(2r,\Z)$ vector bundle over $\cCrg$. This structure makes $\cC$ a special Kahler variety. There are other formulation of special Kahler geometry, some of which will be discussed below, for a systematic account see \cite{Freed:1997dp}.

Finally the complex central charge, $Z_Q$, of the low energy $\cN=2$ supersymmetry algebra of a vacuum in $\cCrg$ acting on the superselection sector of states with charge $Q \in \L$:
\beq\label{ccBPS}
Z_Q := Q^T\s ,
\eeq
where $^T$ indicates transposition. It follows from the $\cN=2$ supersymmetry algebra that $|Z_Q|$ is a lower bound on the mass of any state with charge $Q$.

\subsection{Special Kahler stratification}\label{sec:stra}

The singular locus $\cSb$ plays a central role in our analysis, here we will provide a quick summary of its rich structure, for a more detailed discussion see \cite{Argyres:2020wmq}. It is well-known that the singular locus $\cSb$ is a stratified space with closed subsets of increasingly higher complex co-dimensions. But a careful analysis shows that $\cSb$ inherits an even more constraining structure from the ambient space $\cC$, this is called \emph{special Kahler stratification} in \cite{Argyres:2020wmq}. The main idea behind the special Kahler stratification, is that each component of the singular locus $\cSb_i$, starting at complex co-dimension one, inherits a special Kahler structure from $\cC$ and can be therefore seen as a scale invariant CB geometry in its own right with its own set of metric singularities. As it is the case for $\cCrg$, the set of regular points of $\cSb_i$ will be an open set which we will indicate simply as $\cS_i$ and which we will constitute the \emph{strata} of the stratification. Iterating this procedure we can identify components and corresponding strata of higher complex co-dimensions.

As we discussed throughout the paper, each strata naturally supports a theory which in the text we called $\cT_i$. The direction transverse to each $\cS_i$ into $\cC$ are naturally identified, at least locally, with the CB moduli of the low-energy theory $\cT_i$ supported on the stratum. It therefore follows that the rank of $\cT_i$ coincides with the complex co-dimension of the strata. Restricted in particular, as we did in the text, to complex co-dimension one strata, we only encounter rank-1 theories. Our notation in this section is a bit sloppy and maybe confusing at times since we are not keeping track of the increasing complex co-dimension of the strata and the rank of the corresponding theories. We chose to avoid setting up the proper, lengthy, notation since here we don't really use the properties of higher complex co-dimension strata, for a proper discussion see \cite{Argyres:2020wmq}.

%, which naturally generalizes for higher complex co-dimensional strata. The  rank of the low-energy effective theory describing the massless states along a stratum follows from the rank of its lattice of charges of massless states. The result of the careful analysis of \cite{Argyres:2020wmq} is that it exactly coincides with the complex co-dimension of the stratum. %Let's now focus on the complex co-dimension one strata.

The $\C^*$ action acts by restriction on all strata which are all separately scale invariant and therefore are all closed under it. Since the stratification works inductively, all properties that apply to complex co-dimension one strata extend to the higher complex co-dimension ones. This stratification is very reminiscent of the stratification of symplectic singularities  \cite{beauville1999symplectic,kaledin2006symplectic} which applies to Higgs branches of $\cN=2$ SCFTs in four dimensions. The special Kahler stratification is both more constrained and richer.  It is more constrained because the complex dimension of the strata jump precisely by one at each step and a full list of allowed elementary slices is known, while an analogous list remains an open question for symplectic singularities. And it is richer because strata supporting $\U(1)$ gauge theories with massless hypers and trivial Higgs branch are not necessarily special Kahler and a weaker condition applies \cite{Argyres:2020wmq}.

\subsection{Accidental $\U(1)_R$ near strata}\label{sec:U1R}

%\red{Start with a sentence, that there are two different scaling actions bla bla bla, to orient the reader.}

 The $\U(1)_R$ component of the $R$-symmetry of the theory at the superconformal vacuum, acts non-trivially on $\cC$ and for a generic CB vacuum $\bu$ it will be spontaneously broken. An accidental $\U(1)_R$ symmetry arises instead along any strata $\cS_i$ (and along strata of higher co-dimensions), where this low-energy $\U(1)_R$ is identified with the appropriate component of the global symmetry of $\cT_i$. Because of that, near each stratum, we can define a scaling action which is different from the globally defined one and which does not act globally on $\cC$. This new scaling action scales \emph{into} the stratum and it only really acts on the transverse slice. The global scaling action scales instead into \emph{the origin} of the moduli space.

The accidental $\U(1)_R$ symmetry combines with the $\R^+$ of $\cT_i$ giving rise to a $\C^*_i$ action which scales away from the stratum $\cS_i$. Notice that we did not label this new scaling action as $\C^*|_i$ because $\C^*_i$ \emph{is not} the restriction of $\C^*$. For example $\C_i^*$ is not globally defined and the weight of the coordinates $\bu$ under the $\C^*$ and $\C^*_i$ are in general radically different. The scaling dimension $\D^{\rm sing}_i$ in \eqref{Dsing} is defined with respect to the global $\C^*$ action while the scaling dimension of the CB of the low-energy theories $\cT_i$, $\D_i$ see \eqref{Di}, is defined with respect to $\C^*_i$.

\bibliographystyle{JHEP}

\end{document}